\documentstyle[prd,aps,preprint,psfig,floats]{revtex}
\begin{document}
\newcommand{\gsim}{ \mathop{}_{\textstyle \sim}^{\textstyle >} }
\newcommand{\lsim}{ \mathop{}_{\textstyle \sim}^{\textstyle <} }
\newcommand{\vev}[1]{ \left\langle {#1} \right\rangle }
\newcommand{\eV}{~\mbox{eV}}
\newcommand{\keV}{~\mbox{keV}}
\newcommand{\MeV}{~\mbox{MeV}}
\newcommand{\GeV}{~\mbox{GeV}}
\newcommand{\TeV}{~\mbox{TeV}}
\newcommand{\for}{~~\mbox{for}~}
\newcommand{\fn}{~~\mbox{for}~n=1}
\newcommand{\axino}{{\widetilde a}}
\newcommand{\nnlsp}{{\widetilde \chi_L}}
\tighten
\preprint{\begin{tabular}{l}
\hbox to\hsize{\hfill UT-897}\\
\hbox to\hsize{\hfill}\\
\hbox to\hsize{\hfill}\\
\hbox to\hsize{\hfill}
\end{tabular}}
\title{Solving the Gravitino Problem by Axino}
\author{T.~Asaka$^{1,2}$ and T.~Yanagida$^{2,3}$}
\address{$^1$Deutches Elektronen-Synchrotron DESY,
         22603 Hamburg, Germany}
\address{$^2$Department of Physics,  
         University of Tokyo,
         Tokyo 113-0033, Japan}
\address{$^3$Research Center of the Early Universe,
         University of Tokyo,
         Tokyo 113-0033, Japan}
\date{June 19, 2000}

\maketitle
\begin{abstract}
In a large class of supersymmetric (SUSY) axion model the mass of axino
$\axino$ (a fermionic superpartner of the axion) is predicted as
$m_{\axino} \lesssim {\cal O}(1)$ keV.  Thus, the axino is the lightest
SUSY particle (LSP).  We pointed out that such a light axino provides a
natural solution to the gravitino problem, if the gravitino is the next
LSP.  We derive a constraint on the reheating temperature $T_R$ of
inflation, $T_R \lesssim 10^{15}$ GeV for the gravitino mass $m_{3/2}
\simeq 100$ GeV, which is much weaker than that obtained in the minimal
SUSY standard model.
\end{abstract}
\clearpage

\vskip 1cm
%
%

~

The CP violation in QCD is one of the most serious problems in the
standard model.  In spite of continuous effort to solve the strong CP
problem in the last coupled decades, the mechanism proposed by Peccei
and Quinn~\cite{Peccei-Quinn} is still the most attractive one. The
spontaneous breakdown of the Peccei-Quinn symmetry produces a
Nambu-Goldstone boson (called as axion ``$a$'')~\cite{Weinberg-Wilczek}
and the breaking scale $F_a$ is stringently constrained by laboratory
experiments, astrophysics and cosmology as $F_a \simeq
10^{10}$--$10^{12}$ GeV~\cite{Kim}.

Supersymmetric (SUSY) extension of the Peccei-Quinn mechanism
necessarily predicts a fermionic partner of axion,
\footnote{ The axion supermultiplet $\Phi$ can be written by 
$\Phi = \sigma + i a + \sqrt{2} \theta \axino + \theta^2 F_\Phi$, 
where $a$ denotes an axion, $\sigma$ a saxion, and $\axino$ an axino.
}
  so-called axino $\axino$, whose mass is highly dependent of
models~\cite{AxinoMass1,Rajagopal-Turner-Wilczek,AxinoMass2}.  However,
in a large class of SUSY axion models~\cite{AxinoMass2} the mass of
axino is predicted in the region $m_\axino \lesssim {\cal O}(1)$ keV
which is cosmological harmless~\cite{Rajagopal-Turner-Wilczek}.  In
these models the axino $\axino$ is the lightest SUSY particle (LSP) and
the gravitino $\psi_{3/2}$ can decay into a pair of the axion $a$ and
the axino $\axino$.  We point out, in this letter, that the light axino
provides a natural solution to the cosmological gravitino
problem~\cite{GravitinoProblem}, if the gravitino $\psi_{3/2}$ is the
next LSP. We assume that the axino $\axino$ is the LSP of mass
$m_{\axino} \lesssim {\cal O}$(1) keV and the gravitino is the next LSP
of mass $m_{3/2}\simeq 10^2$ GeV throughout this letter.
\footnote{
This possibility was considered in the context of the galaxy 
formation~\cite{Olive-Schramm-Srednicki}.
However, the gravitino is assumed to have a much longer
lifetime than the estimate in Eq.~(\ref{LT32}),
and hence their analysis in not applicable for the present purpose.
Furthermore, cosmological constraints on the SUSY axion model 
discussed in this letter were not investigated there.}

Radiative decays of gravitinos are cosmologically dangerous, since they
take place after the epoch of the big-ban nucleosynthesis (BBN) and
destroy light nuclei synthesized by the BBN.  To avoid this gravitino
problem, the reheating temperature $T_R$ of inflation should be low
enough~\cite{GravitinoProblem}.  It is found in
Ref.~\cite{Holtmann-Kawasaki-Kohri-Moroi} that the reheating temperature
should be $T_R \lesssim 10^6$ GeV for $m_{3/2} \simeq 100$--$500$ GeV
and $T_R \lesssim 10^8$ GeV for $m_{3/2} \simeq $ 500 GeV--1 TeV.  Such
a low reheating temperature involves significant physical implication,
i.e., it excludes some inflation models and/or baryogenesis scenarios.
The relevant example here is the leptogenesis~\cite{Fukugita-Yanagida}
via decays of heavy Majorana neutrinos to account for the baryon
asymmetry of the present universe.  For the case when heavy Majorana
neutrinos are produced by thermal scatterings, which is the most
conventional production mechanism, a successful leptogenesis requires
the cosmic temperature of $10^{10}$ GeV~\cite{Buchmuller-Plumacher},
which leads to the gravitino problem.  One of motivations in this letter
is to solve this problem by relaxing the above constraints on the
reheating temperatures. As shown below, our hypothesis, i.e, the axino
is the LSP of $m_{\axino} \lesssim {\cal O}$(1) keV and the gravitino is
the next LSP of $m_{3/2} \simeq 10^2$ GeV, allows the reheating
temperature of $10^{15}$ GeV which is sufficiently high for the
thermal leptogenesis to work.
\footnote{ Another solution had been
proposed in Ref.~\cite{Bolz-Buchmuller-Plumacher}.  }

In the present model, the main decay of the gravitino $\psi_{3/2}$ is
$\psi_{3/2} \rightarrow \axino + a$ and its lifetime is estimated as
\begin{eqnarray}
 \label{LT32}
 \tau_{3/2} \simeq \frac{ 192 \pi M_\ast^2}{m_{3/2}^3}
 \sim 10^{9} ~\sec~
  \left( \frac{ 10^2 \GeV}{ m_{3/2} }\right)^3 ~,
\end{eqnarray}
where $M_\ast = 2.4 \times 10^{18}$ GeV is the reduced Planck mass.
Since $\axino$ and $a$ have very weak couplings to the ordinary
particles, this gravitino decay does not destroy any BBN light nuclei.
The ratio of the gravitino energy density to the entropy density is
given by~\cite{GravitinoProblem}
\begin{eqnarray}
 \label{AB32}
 \frac{ \rho_{3/2}}{s} 
  \sim
  10^{-9} \GeV ~
  \left( \frac{ m_{\widetilde g} }{ 1 \TeV}\right)^2
  \left( \frac{ T_R }{ 10^{10} \GeV }\right)
  \left( \frac{ m_{3/2}}{10^2 \GeV }\right)^{-1}~,
\end{eqnarray}
where $m_{\widetilde g}$ is the gluino mass.  The ratio $\rho_{3/2}/s$
in Eq.~(\ref{AB32}) should be smaller than about $10^{-4}$ GeV.
Otherwise, this extra energy density raises the expansion rate of the
universe at the BBN epoch and leads to overproduction of $^4$He.  This
gives an upper bound on the reheating temperature as $T_R \lesssim
10^{15}$ GeV for $m_{3/2}=10^2$ GeV.
\footnote{ 
A similar constraint on $T_R$ was obtained for the lighter gravitino of
mass $\sim 100$ MeV in Ref.~\cite{Olive-Schramm-Srednicki} from their
scenario of the structure formation.  However, our condition from
Eq.~(\ref{AB32}) leads to a more stringent constraint on the reheating
temperature $T_R \lesssim 10^{12}$ GeV for such a light gravitino.
}

On the other hand, the lightest SUSY particle $\nnlsp$ next to the
gravitino decays into a gravitino emitting photons. If this is only the
decay mode, the energetic photons destroy the light nuclei and cause a
serious problem in the BBN~\cite{GravitinoProblem}. This is because the
decay takes place soon after the BBN ends.  However, in the present
model such a particle can decay mainly into an axino and a photon,%
\footnote{
$\nnlsp$ is assumed to be mainly composed of the photino
$\widetilde \gamma$.  }
   and its decay lifetime is~\cite{KMN-N}%
\footnote{
If the $R$-parity is broken, $\nnlsp$ can decay into the ordinary
light particles avoiding the problem in the BBN.
}
\begin{eqnarray}
 \tau_{\widetilde \chi_1} \sim 10^{-3} ~ \sec ~
  \left( \frac{ F_a }{ 10^{11} \GeV }\right)^2
  \left( \frac{ 10^2 \GeV }{ m_{\widetilde \chi_1} } \right)^3~.
\end{eqnarray}
That is, it decays much before the BBN starts and hence
there is no problem at all.

Thus, the problem we must discuss below is whether the SUSY axion model
with the light axino is cosmologically safe or not.  First, we discuss
cosmological abundance of $\axino$, especially, the overclosure
problem of the LSP axino.  

Let us discuss possible production mechanisms of the axinos $\axino$.
In the early universe the axinos are produced in the thermal equilibrium
though the reactions like $q \overline{q} \leftrightarrow \axino
\widetilde g $, and it decouples from the thermal bath at the cosmic
temperature~\cite{Rajagopal-Turner-Wilczek}
\begin{eqnarray}
 \label{Td}
 T_d \sim 10^{9} \GeV
  \left( \frac{ F_{a} }{ 10^{11} \GeV} \right)^2 ~.
\end{eqnarray}
If the reheating temperature of inflation is higher than this decoupling
temperature ($T_{R} \gg T_d$), the yield of the axino $Y_{\axino}$
($Y_{\axino} \equiv n_\axino / s$ with the axino number density
$n_\axino$ and the entropy density $s$) is estimated as
\begin{eqnarray}
 \label{Yaxino1}
 Y_{\axino} \equiv \frac{ n_\axino }{s}
  \sim 10^{-3} ~.
\end{eqnarray}
On the other hand, for $T_{R} \ll T_d$,
the yield of the axino is given by
\begin{eqnarray}
 \label{Yaxino2}
 Y_{\axino} 
  \sim 10^{-3} 
  \left(  \frac{ T_{R}}{T_d} \right)~.
\end{eqnarray}
For the case of the stable LSP axino, the present energy density 
of the axino may exceed the critical density of the present
universe in some parameter regions.
The density parameter of the axino is 
\begin{eqnarray}
 \Omega_\axino = \frac{ m_\axino Y_\axino}{ \rho_c/s_0}~,
\end{eqnarray}
where $\rho_c$ is the critical density and $s_0$ denotes
the total entropy density of the present universe
($\rho_c/s_0 = 3.6 \times 10^{-9} h^{2}\GeV$ with the Hubble parameter $h$
in unit of 100 km/sec/Mpc.).
From Eq.~(\ref{Yaxino1}) we find
\begin{eqnarray}
 \Omega_\axino h^2 
  \simeq 5.8 \times 10^{5} 
  \left( \frac{ m_\axino}{1 \GeV}\right) ~.
\end{eqnarray}
Therefore, the non-overclosure limit $\Omega_\axino h^2 \lesssim 1$ gives
the upper bound on the axino mass as~\cite{Rajagopal-Turner-Wilczek}
\begin{eqnarray}
 \label{UBmaxino}
 m_\axino \lesssim 2 \keV ~.
\end{eqnarray}
On the other hand, if $T_{R} < T_d$, we find
that the upper bound (\ref{UBmaxino}) is relaxed as
\begin{eqnarray}
 \label{UBmaxino2}
  m_\axino \lesssim 2 \keV \left( \frac{ T_d }{T_{R} }\right)~.
\end{eqnarray}
It should be noted here that we have a large class of SUSY axion 
models~\cite{AxinoMass2} with such a light axino, as mentioned 
in the introduction.

The axinos are also produced by the decays of gravitino and $\nnlsp$.
However, they are safely neglected because the axino mass 
should be small enough to satisfy the condition 
Eq.~(\ref{UBmaxino}) or (\ref{UBmaxino2}).

Next, we turn to discuss the cosmological problem associated with the
saxion $\sigma$.  The saxions are also produced through the thermal
scattering processes as well as the axinos, and its decoupling
temperature is also given by Eq.~(\ref{Td}). Therefore, the yield of the
saxion is estimated as
\begin{eqnarray}
 Y_\sigma \sim 
  \left\{
   \begin{array}{l l }
    10^{-3}  & \for T_R \gg T_d
     \\
    10^{-3} \left( \frac{ T_R}{ T_d} \right) & \for T_R \ll T_d
   \end{array}
\right. ~.
\end{eqnarray}
Then, the ratio of the saxion energy density to the entropy density 
is given by
\begin{eqnarray}
 \label{ABsTH}
 \frac{\rho_\sigma}{s} \sim 
  \left\{
   \begin{array}{l l }
    10^{-3} m_\sigma & \for T_R \gg T_d
     \\
    10^{-3} m_\sigma \left( \frac{ T_R}{ T_d} \right) & \for T_R \ll T_d
   \end{array}
\right. ~.
\end{eqnarray}

Notice that the saxion mass is comparable to the gravitino mass
($m_\sigma \sim m_{3/2}$).  For $m_\sigma \simeq 10^2$ GeV the saxions
dominate the energy density of the universe after the cosmic temperature
$T$ cools down to $\sim 100$ MeV.  However, the saxion is not stable.
The relevant decay channels are $\sigma
\rightarrow 2 g$ and $\rightarrow 2 a$, and their decay rates are
estimated as
\begin{eqnarray}
 &&\Gamma_{\sigma \rightarrow 2 g } =
  \frac{ \alpha_s^2 }{32 \pi^3}
  \frac{ m_\sigma^3}{ F_a^2} ~,
\\
 &&\Gamma_{\sigma \rightarrow 2 a } =
  \frac{ C }{32 \pi} 
  \frac{ m_\sigma^3}{ F_a^2} ~,
\end{eqnarray}
where $C$ is the constant of $C \lesssim 1$.  Since the constant $C$
depends on the model for the $U(1)_{PQ}$ symmetry breaking, we take
it as a free parameter.
When the constant $C$ is large enough as
\begin{eqnarray}
 C \gtrsim 0.8 
  \left( \frac{ 10^2 \GeV}{m_\sigma} \right)
  \left( \frac{ F_a }{10^{11} \GeV} \right)^2,
\end{eqnarray}
the saxions decays mainly into axions much before 
they dominate the energy of the universe and 
the produced axions are small enough.
On the other hand, when $C$ becomes smaller than this critical value,
the $\sigma \rightarrow 2 a$ decay channel
should be suppressed enough, otherwise the extra energy density 
of the produced axions at the cosmic temperature $T \sim 1$ MeV spoils the 
success of the BBN.  In order to avoid this difficulty,
the branching ratio of the saxion decay into two axions
should be smaller than about 0.1, i.e., $C \lesssim 10^{-4}$.
In this case the saxions dominate the universe before they decay,
the universe is reheated again by the saxion decay.  The reheating
temperature $T_\sigma$ is estimated as
\begin{eqnarray}
 T_\sigma
 \sim
 56\MeV~
 \left( \frac{ m_\sigma}{10^2 \GeV}\right)^{3/2}
 \left( \frac{ 10^{11} \GeV}{ F_a}\right)~.
\end{eqnarray}
Therefore, the saxion decay completes before the BBN starts.  If the
Peccei-Quinn breaking scale is large as $F_a \sim 10^{12}$ GeV, the
saxion decay increases the entropy density of the universe by the factor
$\Delta$~\cite{Kim-Lyth-IHYY}
\footnote{ If there is an entropy production after the QCD phase
transition, the upper bound on $F_a$ is raised up above
$F_a \simeq 10^{12}$ GeV \cite{Kawasaki-Moroi-Yanagida}.  }
\begin{eqnarray}
 \label{EP}
 \Delta \sim 24
  \left( \frac{10^2 \GeV}{m_\sigma}\right)^{1/2}
  \left( \frac{ F_a }{ 10^{12} \GeV }\right)~.
\end{eqnarray}
However, there is no entropy production by the saxion decay
for the case of $F_a \simeq 10^{10}$--$10^{11}$ GeV.
For $T_R \ll T_d$, the entropy production rate $\Delta$
is suppressed by the factor $(T_R/T_d)$ and no entropy production
takes place when $T_R/T_d \lesssim 0.04$.

Furthermore, it should be noted that the saxion may be produced
effectively in the form of the coherent oscillation after the inflation.
We assume here that the supergravity effects induce positive mass
squared for Peccei-Quinn scalar fields
\footnote{
Peccei-Quinn fields are scalar fields responsible for the Peccei-Quinn
symmetry breaking.
}
   of order of $H_I$ during the inflation ($H_I$ 
denotes the Hubble parameter for the inflation).
Thus, it is quite natural that the Peccei-Quinn symmetry is restored
during the inflation if $H_I > F_a$.
\footnote{ Since the $U(1)_{PQ}$
symmetry is restored during the inflation, there is no massless mode and
hence no isocurvature fluctuation is generated~\cite{Asaka-Yamaguchi}.  }
The Peccei-Quinn symmetry breaking occurs when the Hubble parameter becomes
comparable to the scale $F_a$, and the coherent oscillation starts.
Then, the ratio of the energy density of the coherent oscillation
$\rho_{\rm osc}$ to entropy density for $T \ll T_R$ is 
estimated as
\begin{eqnarray}
 \frac{ \rho_{\rm osc} }{ s}
  &=& \frac{ 1}{8}
  \frac{ T_R F_a^2}{ M_G^2}~,
\nonumber \\
  &\simeq&
   10^{-6} \GeV~
   \left( \frac{ T_R }{ 10^{10} \GeV } \right)
   \left( \frac{ F_a }{10^{11} \GeV}\right)^2~.
\end{eqnarray}
Comparing it with the energy density of the saxion produced by the
thermal scatterings in Eq.~(\ref{ABsTH}), we can safely neglect the
energy density of the oscillation if $T_R \lesssim 10^{15}$ GeV
for $F_a \simeq 10^{11}$ GeV.

Finally, we should comment on the axion domain walls.
We have assumed that the $U(1)_{PQ}$ symmetry is restored
during the inflation, and hence the axion domain walls 
might be formed after the inflation ends.
However, this can be easily evaded by adopting a hadronic 
axion model~\cite{HadronicAxion} with the domain wall number 
$N_{DW}=1$~\cite{Kim}.

In this letter we have pointed out that the cosmological gravitino
problem can be solved by the SUSY axion model which is the most natural
solution to the strong CP problem.  In the present model, the axino is
the LSP and the gravitino is the next LSP.  The gravitino decays into a
pair of the axino and the axion eluding the photo-dissociation
constraint on the reheating temperature $T_R \lesssim 10^6$--$10^8$ GeV which
was obtained in the minimal SUSY standard model with an unstable
gravitino of $m_{3/2} \simeq 100$ GeV--1
TeV~\cite{Holtmann-Kawasaki-Kohri-Moroi}.  
We find that the reheating temperature can be high as $T_R \simeq
10^{15}$ GeV for $m_{3/2} \simeq 10^2$ GeV.
Therefore, the present model
makes thermal leptogenesis scenarios~\cite{Buchmuller-Plumacher} to work
well without the cosmological gravitino problem.%
\footnote{
The primordial lepton (baryon) asymmetry is not diluted away by 
the saxion decay,
since we have only a small entropy production as shown in Eq.~(\ref{EP}).
}

%
%
\section*{Acknowledgements}
This work was partially supported by the Japan Society for the
Promotion of Science (T.A).

\end{document}